\documentstyle[preprint,prb,tighten,aps,epsf]{revtex}
\begin{document}
\draft
\title{Electron Correlations in Molecular Systems}
\author { Bogdan R. Bu{\l}ka\cite{ebb}}
\address{Institute of Molecular Physics,
Polish Academy of Sciences,   \\
ul. Smoluchowskiego 17, 60-179 Pozna\'n, Poland}
\date{Received \hspace{5mm} } 
\date{\today} 
\maketitle
\begin{abstract}
  A short review of correlated electrons in molecular systems has been
  performed. Main attention has been focussed on ET salts, which are
  the $d$=2 systems. They show the Mott transition in high
  temperatures and the transition from the antiferromagnetic to the
  superconducting phase in low temperatures, under a (chemical)
  pressure. Physical properties (the electrical resistivity, the
  specific heat, the magnetic susceptibility, the photoemission
  spectra, the optical conductivity) of ET salts have been compared
  with those ones in other strongly correlated systems.  The optical
  conductivity is described in the framework of the Hubbard model,
  with a low frequency peak as an evidence for the Abrikosov-Suhl
  resonance.
\end{abstract}

\section{INTRODUCTION}

Recently we become more aware that interactions between electrons play
an important role in many materials. Electron correlations in space
and time, their fluctuations can be strong and therefore many physical
properties are different than those for conventional metals described
by a free electron picture. A list of materials with strongly
correlated electrons, in which one can observe such differences,
become longer and longer. Different types of materials are classified
to the same class if kinetics of electrons is geometrically restricted
to a {\it d} dimensional space. Despite their differences in a
chemical composition, the materials of each class show similarities in
many physical characteristics. Let us mention some of them.
\cite{a1,a2,a3}. Among {\it d}=3 correlated electron systems are:
magnetic transition metals with 3d electrons, materials with the Mott
type of the metal-insulator transition (e.g. V$_2$O$_3$), and heavy
fermion systems (as UPt$_3$, URu$_2$Si$_2$, CeAl$_3$). If one extends
the list to strongly correlated fermions, so liquid $^3$He should be
placed there as well. With lowering dimensionality of the strongly
correlated system their physical properties become more unusual, as
those in high temperature superconductors with weakly coupled layer of
conducting electrons. To the $d$=2 system one can classified also
organic conductors and superconductors (e.g. ET salts, where ET
$\equiv$ BEDT-TTF is bis(ethylenodithio)tetrathia-fulvalene),
electronic systems showing a fractional quantum Hall effect and
magnetic metallic multilayers. The so-called spin ladders are systems
of localized spins and itinerant electrons, which dimensionality are
closer to $d$=1 than $d$=2. Conducting polymers, organic conductors
and organic superconductors are well known strongly correlated $d$=1
materials. There are also in this class inorganic compounds with a
charge density wave (CDW) and quantum wires as well edges of the $d$=2
systems in the quantum Hall effect. Due to developments in
mircrotechnology we have also the $d$=0 systems as quantum dots and
single electron transistors made of different materials.

Our knowledge on strongly correlated electrons increased significantly
in recent 10 years. Enormous interest in high temperature
superconductors gave also a new impact for studies of $d$=2 molecular
systems. The aim of this paper is to review some of most interesting
achievements in investigations of organic superconductors from the ET
family.

The paper is organized in the following. We begin with a model
description of some simple electron correlated systems. We show that
the $d$=2 extended Hubbard model has the ground state with the
antiferromagnetic phase (AF), the CDW or the superconducting phase of
the s- and d- type pairing depending on interactions parameters. In
high temperatures the system can be a metal or a Mott insulator. The
physical quantities of the metallic phase can be derived within the
Fermi liquid theory. The main part is an overview the properties of
(ET)$_2$X, where X= with X=Cu(NCS)$_2$, Cu[N(CN)$_2$]Br and
Cu[N(CN)$_2$]Cl. The physical properties of the conducting (ET)$_2$X
salts, as the electrical conductivity, the specific heat and the
magnetic susceptibility, will be analyzed within the framework of the
Fermi liquid theory. In this series of compounds the AF ordering and
superconductivity of the d-type was found as well as the
metal-insulator transition of the Mott type in high temperatures. We
show photoemission studies of electronic structure and optical
conductivity measurements, which indicate on strong correlations in
these materials.

\section{Model of correlated electrons and its basic properties}

The hydrogen molecule H$_2$ is the simplest system with electron
correlations. Let us remind that to describe a two electron state
Slater postulated the wave function in the form\cite{a4}
\begin{equation}\label{e1}
\Psi_s(r_1,r_2) = \frac{1}{2} [\Phi_1(r_1)\Phi_1(r_2)+\Phi_1(r_1)\Phi_2(r_2)+\Phi_2(r_1)\Phi_1(r_2) +\Phi_2(r_1)\Phi_2(r_2)] (\alpha_1\beta_2 - \alpha_2\beta_1)\;,
\end{equation}
where $\Phi(r_m)$ is the single electron wave function for the $m$-th
electron on the $i$-th atom and $\alpha_m$, $\beta_m$ are spinors
describing spins up and down, respectively. The wave function {e1}
corresponds to free (uncorrelated) electrons. It is seen that in this
basis the average for two electrons on a given hydrogen atom
$<n_{1\uparrow}n_{1\downarrow}> = <n_{1\uparrow}><n_{1\downarrow}> =
1/4$. The Heitler-London wave function: function in the form\cite{a4}
\begin{equation}\label{e2}
\Psi_s(r_1,r_2) = \frac{1}{2} [\Phi_1(r_1)\Phi_2(r_2)+\Phi_2(r_1)\Phi_1(r_2)] (\alpha_1\beta_2 - \alpha_2\beta_1)\;,
\end{equation}
corresponds to the extreme strong correlations. In this case the
average $<n_{1\uparrow}n_{1\downarrow}> =0$. It is well
known~\cite{a4} that the exact solution for H$_2$ shows finite
correlations with a wave function being the linear combination
(\ref{e1}) and (\ref{e2}).

In the case of many electrons an exact solution is difficult to find.
The simplest approximation used in such the case is the Hartree-Fock
(HF) one
\begin{eqnarray}\label{e3}
<a_{i\sigma}^{\dag}a_{k\sigma'}^{\dag}a_{l\sigma'}a_{j\sigma}>\approx 
\left\{\begin{array}{lll}
<a_{i\sigma}^{\dag}a_{j\sigma}><a_{k\sigma'}^{\dag}a_{l\sigma'}>&
\mbox{for}&\sigma=\sigma'\;,\\ 
<a_{i\sigma}^{\dag}a_{j\sigma}><a_{k\sigma}^{\dag}a_{l\sigma}>
-<a_{i\sigma}^{\dag}a_{l\sigma}><a_{k\sigma}^{\dag}a_{j\sigma}>&
\mbox{for}&\sigma\ne \sigma'\;.\end{array}\right.
\end{eqnarray}
Here, $a_{i\sigma}$ denotes the annihilation operator of an electron
with spin s at the i-th site of the lattice. This approximation for
H$_2$ corresponds to the solution with the Slater wave function
(\ref{e1}) and $<n_{1\uparrow}n_{1\downarrow}> =
<n_{1\uparrow}><n_{1\downarrow}> = 1/4$. Electron correlations are
neglected in the HF approximation.  The exact results for the ground
state energy can be significantly different from that one obtained
within the HF approximation. For example, for a polyethylene chain the
correlation energy, defined as
$E_{corr}=E_{ground}^{exact}-E_{ground}^{HF}$ , is -3.97
eV/monomer.~\cite{a2}. A simplest conducting polymer is polyacetylene
(CH)$_x$. It is known that the Su-Schrieffer-Heeger model~\cite{a5} is
not sufficient to describe quantitatively the physical properties of
the $d$=1 chains of polyacetylene. It is needed to extend the model
including local interactions of electrons. The Hamiltonian for
(CH)$_x$ is given by
\begin{equation}\label{e4}
H=-\sum_{<i,j>,\sigma}[t+(-1)^{i}2\alpha\xi]c_{i\sigma}^{\dag}c_{j\sigma}
+U\sum_{i}n_{i\uparrow}n_{i\downarrow}+V\sum_{<i,j>}n_in_j+2NK\xi^2\;,
\end{equation}
where the parameters determined in ab initio calculations
are~\cite{a2}: the hopping integral $t$ = 2.5 eV, the electron-phonon
coupling $\alpha$ = 40 meV/pm, the elastic constant $K$= 3.9 meV/pm2,
the onsite Coulomb integral $U$= 11.5 eV, the intersite Coulomb
integral $V$= 2.4 eV. The first term describes the kinetics of
electrons in presence of the Peierls distortion $\xi$ of the lattice,
the last term is the elastic energy of the distorted lattice, the
second and the third term correspond to the onsite and the intersite
interaction of electrons. The electron-electron interactions in
polyacteylene are moderately strong $U \approx W$ ($W=4t$ is the width
of the electronic band).

The tight binding Hamiltonian for interacting electrons on the lattice
was first formulated and considered by Hubbard.~\cite{a6}. It includes
the kinetic term and the onsite interactions. The model described by
(\ref{e4}) is called the extended Hubbard model, as it has additional
terms. The extended Hubbard model with different intersite
interactions has been recently intensively analyzed in hope to find an
appropriate description of high temperature superconductors. Many
physical properties in these materials indicate on strong electron
correlations. The phase diagram oxide cuprates shows the AF ordering
as well as the supercondcuting phase.~\cite{a7}. Their relative
stability depends on electron doping. Our studies~\cite{a8} of the
stability of the ground state for the half-filled band of the $d$=2
extended Hubbard model performed within the HF approximation showed
the AF ordering, if the onsite repulsion term dominates ($U> |V|$) -
see Fig.1. For lower values of $U$ the ground state stable solution is
with CDW or superconducting, depending on the sign of the intersite
coupling $V$. The superconducting state is the d-type symmetry for
small negative values of $V$ and a mixed s- and d-type, for larger
couplings $|V|$.

Studies in finite temperatures are extremely difficult because one has
to take into account, apart from correlations, various types of
fluctuations in space and time. To overcome the problem one can
analyze either in the limiting cases, of weak $U\ll t$ and strong
onsite repulsion $U\gg t$, by means of the perturbation approach or
using slave bosons. We performed the slave boson studies~\cite{a9} of
the stability of the ground state with respect to the normal phase,
which can be either metallic one or a Mott insulator, depending on the
value of $U$.  The slave boson approach includes electron correlations
and give reliable results in all range of $U$. The condensation energy
and the critical temperature $T_c$ for the AF phase increases with
$U$, in the weak coupling limit, has a wide maximum at $U\approx W$
and decreases for large $U$.

The Mott transition to an insulating phase occurs in high temperatures
only for the half-filled band. In the other cases the high temperature
phase of a strongly correlated electron system is metallic, but its
properties are different than those in conventional metals. The $d$=1
system of correlated electrons is described by the {\it Luttinger
  liquid} and in higher dimension ($d\ge 2$) by the {\it Fermi
  liquid}.~\cite{a2} The concept of the Fermi liquid relies on the
notion of quasiparticles, which have an energy spectrum ek and an
effective relaxation time $t$. The Green's function, describing
dynamics of quasiparticles, is
\begin{equation}\label{e5}
G(k,\omega)= \frac{1}{\omega-\varepsilon_k^0+\Sigma(k,\omega)}
=\frac{Z}{\omega-\varepsilon_k+i\hbar\tau^{-1}(\omega)}+G_{incoh}(k,\omega)\;,
\end{equation}
where $\Sigma(k,\omega)$ is a self-energy, $Z^{-1} = 1-\partial
\Sigma(k,\omega)/\partial \omega|_{\omega=0} = m^*/m$ is a
renormalization constant, $G_{incoh}$ is an incoherent part of the
Green's function.

The function (\ref{e5}) can be used in derivations of many physical
quantities, as the specific heat, different susceptibilities, and
transport properties.~\cite{a2} They involve low-energy excitations of
the Fermi liquid. The excited electrons are close to the Fermi energy
$\varepsilon_F$ and the effective relaxation time is
\begin{equation}\label{e6}
\tau^{-1}=a(\omega-\varepsilon_F)^2+bT^2\;.
\end{equation}
The electron mean free path due to electron-electron interaction is
then $l_{el-el} = v_F\tau$, where $v_F$ is the velocity at the Fermi
energy. In low temperatures the electrical resistivity is therefore
\begin{equation}\label{e7}
\rho=\rho_0+AT^2\;,
\end{equation}
where $\rho_0$ is the residual resistivity. The specific heat is, in
low temperatures, given by
\begin{equation}\label{e8}
C=\gamma T+\delta T^3\ln T\;,
\end{equation}
where $\gamma=k_B^2m^*k_F/3$ and $k_F$ is the Fermi wave vector. The
Sommerfeld coefficient $\gamma$ as well as the effective mass $m^*$
are enhanced due to correlations. The magnetic susceptibility is
expressed
\begin{equation}\label{e9}
\chi=\chi_0\frac{m^*/m}{S}\;,
\end{equation}
where $\chi_0$ is the magnetic susceptibility of free electrons and
$S$ is the Stoner enhancement factor, which plays an important role in
magnetic properties of transition metals.

\section{Electron correlations in ET family}

Recently Kanoda studied~\cite{a10} series of (ET)$_2$X salts with
X=Cu(NCS)$_2$, Cu[N(CN)$_2$]Br and Cu[N(CN)$_2$]Cl as well as their
deuterated forms. All these salts are isostructural; they are in the
$\kappa$ structure. From the temperature dependences of electrical
resistivity, magnetic susceptibility, NMR as well as specific heat
measurements Kanoda proposed~\cite{a10} the schematic phase diagram
presented in Fig.2. The specific heat data for (ET)$_2$Cu[N(CN)$_2$]Br
showed~\cite{a11} that a low temperature phase ($T<$13K) is
superconducting. Moreover, the dependence of the Sommerfeld
coefficient $\gamma$ in magnetic field is $\gamma(H) = A(H+H^*)^1/2$,
what indicates on the superconducting phase with line of
nodes.~\cite{a12}. In very low temperatures ($T<$1K) the specific
heat~\cite{a11} exhibits an activated character suggesting on the
s-type pairing as well. It means that the superconducting phase is the
mixed s+d-type, with the d-component as a dominant one. The magnetic
susceptibility data for (ET)$_2$Cu[N(CN)$_2$]Cl showed~\cite{a13} at
$T < T_N = 26\div27$K an anisotropy of the susceptibility for a field
perpendicular and parallel to magnetization - typical for the AF
ordering (a small canting was also found in a direction parallel to
the conducting layers). In the phase diagram, shown in Fig.2,
(ET)$_2$Cu(NCS)$_2$ and (ET)$_2$Cu[N(CN)$_2$]Br are on the metallic
side, (ET)$_2$Cu[N(CN)$_2$]Cl on the insulating side, and deuterated
(ET)$_2$Cu[N(CN)$_2$]Br just on the borderline. An applied pressure
decreases the critical temperature $T_c$.~\cite{a10} The electronic
bandwidth $W$ increases with pressure and the ratio $U_{eff} / W$
decreases (as the effective intradimer Coulomb repulsion $U_{eff}$ is
less sensible).

The phase diagram for the ET family (Fig.2) has a similar shape as
that one for cuprates ~\cite{a7}, which has also the AF ordering close
to the superconducting phase. However, in the present case the
critical temperature is plotted vs. the ratio $U_{eff} / W$, whereas
in cuprates it is a function of the electron concentration $n$. The
electron concentration in the ET salts is always $n$ =1 (per a dimmer,
which is considered as an effective lattice site). The ratio $U_{eff}
/ W$ is estimated to be close to unity.~\cite{a10} The diagram in
Fig.2 is in agreement with the diagram of the ground state for the
extended Hubbard model presented in Fig.1. If the considered system
has the AF ordering, then with increasing pressure the value $U/t$
decreases, while $V/t$ remains unchanged. Thus, for $V/t <$0 the
system undergoes from the AF to the superconducting phase, which can
be the s+d-type if the coupling is strong enough.

Photoemission spectroscopy of electrons is a direct measurement of
dynamics of electrons and the Green's function (\ref{e5}). In the
experiment one measures a spectral function $A(k,\omega)$ (or
$B(k,\omega)$) by a direct (or an inverse) photoemission effect, which
are expressed by
\begin{equation}\label{e10}
\mbox{Im} G(k,\omega) = -\pi A(k,\omega) \Theta(\omega-\varepsilon_F) + \pi B(k,\omega) \Theta(\varepsilon_F-\omega)\;.
\end{equation}
High-resolution photoemission studies were performed on the $\kappa$-(ET)$_2$X salts by Sekiyama et al. [\onlinecite{a14}]. The photoemission spectra showed the electronic structure with many HOMO bands. In order to analyze a HOMO band with a smallest binding energy they extracted other bands from the photoemission spectra.  In all investigated ET salts, the intensity is suppressed near eF, similarly as it was observed in various oxides,~\cite{a15} indicating on strong correlations of electrons.  The spectra deviate from the HF calculations of the electronic stucture. To explain the suppression of the intensity near $\varepsilon_F$, it was needed to assume a moderate mass renormalization for (ET)2Cu[N(CN)2]Cl and a stronger renormalization for (ET)$_2$Cu(NCS)$_2$ and (ET)$_2$Cu[N(CN)$_2$]Br. It is contrast with the diagram proposed by Kanoda~\cite{a10}, in which correlations are strongest in (ET)$_2$Cu[N(CN)$_2$]Cl. 

According the Fermi liquid approach a low temperature resistivty
should exhibit $T^2$ dependence [Eq.(\ref{e7})] and its coefficient
$A$ has to be coupled with the coefficient $\gamma$ in the specific
heat. Fig.3 collects the data for heavy fermion compounds, A15
materials, transition metals, fullerenes and organic
conductors.~\cite{a16,a17,a18,a19} The ratio $A/\gamma^2 = 1\times
10^{-11} [\Omega cm(mol K/mJ)^2]$ is universal value for heavy fermion
systems (solid line in Fig.3) and $A/\gamma^2 = 4\times 10^{-13}$ for
transition metals (dashed line). Such two values are determined by
Miyake et al. [\onlinecite{a16}] as limiting ones for the Fermi
liquid. Their approach~\cite{a16} was based on a phenomenological
analysis of a frequency dependence of the self-energy
$\Sigma(k,\omega)$ of the Green's function and its influence on $A$
and $\gamma$. The value $A/\gamma^2 = 1\times 10^{-11}$ is obtained
for moderate many-body correlations, i.e. when
$|\partial$Re$\Sigma(\omega)/\partial\omega|>1$ . If a large effective
mass $m^*$ of electrons is due to the properties of a single-body-band
and $|\partial$Re$\Sigma(\omega)/\partial\omega|<1$, then the ratio
$A/\gamma^2$ is smaller and close to $4\times 10^{-13}$.

The data for the organic compounds (full dots) lie in Fig.3 much above
the solid line, the ratio $A/\gamma^2$ is much higher than the upper
limit $1\times 10^{-11}$. There are some explanations of this fact. A
mechanism for $T^2$ dependence of the resistivity may be phononic (or
libronic). In the presence of disorder and strong electron-phonon
coupling the temperature dependence of the resistivity can be (after
Gurvitch~\cite{a20}) proportional to $T^2$. Other possibility is an
error in determination of the coefficient $A$. In Fig.3 the diamonds
represent the data for Sr$_2$RuO$_4$,~\cite{a18} which is a $d$=2
system with highly conducting planes. Precise measurements of the
resistivity performed parallel and perpendicular to the conducting
plane gave~\cite{a18} two different values of $A$ (the points are
denoted as $\parallel$ and $\perp$ in Fig.3, respectively). One may
suspect that the presented data for the organic conductors were
performed no precisely along the conducting plane (or chain) and they
contained a perpendicular component as well. Precise measurements of
the resistivity would show an anisotropy of $A$ and new points in the
plot $A$ vs. $\gamma$ would be placed lower than the present ones.

In Fig.4 the dependence of the Pauli susceptibility vs. the
coefficient $\gamma$ is presented for heavy fermion compounds, to
which the data for the organic conductors and the metal oxides have
been added. The points are close to the solid line corresponding to
the Wilson ratio $R=\frac{\chi/\chi_0}{\gamma/\gamma_0}=2$. It means
that the organic conductors differ from single-body-band systems with
$R$ = 1, and correlations play an important role.

An analysis of optical properties of correlated metals is a much
difficult task. The optical conductivity is expressed in terms of a
current-current correlator $<[j, j]>$ as
\begin{equation}\label{e11}
\sigma(\omega)=-\frac{1}{\omega}\mbox{Im}<[j,j]>\;.
\end{equation}
The function $<[j, j]>$ describes dynamics an electron and a hole, and therefore, it is called as a two-particle Green's function. It is more difficult to determine the two-particle Green's function than the one-particle Green's function (\ref{e5}). 

The formula (\ref{e11}) can be rewritten in the form~\cite{a21}
\begin{equation}\label{e12}
\sigma(\omega)=-\frac{<T_x>}{i\omega}-\frac{1}{i\omega}\sum_{n>0}\frac{|<n|J_{px}|0>|^2}{\omega-(E_n-E_0)+i\hbar\tau^-1}\;.
\end{equation}
Here, $T_x$ and $J_{px}$ denote the kinetic energy operator and the
paramagnetic current operator along the current direction $x$,
respectively. The sum is over all excited states n of the system. For
a system with a discrete translation invariance (e.g. such as the
Hubbard model) and for the relaxation time $\tau$ independent on the
energy, one gets
\begin{equation}\label{e13}
\sigma(\omega)=-\frac{e^2}{Vd}\frac{D}{\pi}\frac{\tau}{1+\omega^2\tau^2}+\sigma_{incoh}(\omega)\;,
\end{equation}
where $\sigma_{incoh}$ is an incoherent part of the conductivity, $D =
\omega_p^{*2} /4\pi$ is a Drude weight and $\omega_p^*$ is a
renormalized plasma frequency. In the limit $\omega\to 0$ one obtains
from (\ref{e12})
\begin{equation}\label{e14}
\frac{\omega_p^{*2}}{8}=-\frac{\pi}{2}<T_x>- \pi \sum_{n>0}\frac{|<n|J_{px}|0>|^2}{(E_n-E_0)}\;.
\end{equation}
The integral 
 \begin{equation}\label{e15}
\frac{\omega_p^{2}}{8}\equiv\int d\omega \sigma(\omega) =-\frac{\pi}{2}<T_x>
\end{equation}                                                     
is the total oscillator strength (or the spectral weight). For
noninteracting electrons the second term in (\ref{e14}) vanish in the
thermodynamic limit and therefore, $\omega_p^{*2}=\omega_p^2$ (as it
is in the Drude model). For the model of interacting electrons the sum
in (\ref{e14}) becomes nonzero and thus~\cite{a21}
 \begin{equation}\label{e16}
\omega_p^{*2}\le\omega_p^2\;.
\end{equation} 

For all organic conductors the total oscillator strength $\omega_p^2$
, calculated by integration of optical spectra, is significantly
smaller than $\omega_p^{*2}$, determined from the band
edge.~\cite{a17,a22} For example, in (ET)$_2$Cu(NCS)$_2$:
$\omega_p^{*2} = 7.6\times10^{-7}$ cm$^{-2}$ and $\omega_p^{*2}=
4.9\times10^{-7}$ cm$^{-2}$ (for polarization of light parallel to the
b-axis).~\cite{a17} An explanation of the controversy lies in
oversimplified interpretation of optical spectra. Jacobsen~\cite{a22}
and others used a single band model to description of measured
spectra. They fitted the spectrum near the band edge to the Drude
model [Eq.(\ref{e13})]. The optical conductivity $\sigma(\omega)$ of
organic conductors has a charge transfer band located at
2000$\div$4000 cm$^{-1}$ and a peak at low frequencies. Such the shape
show also the optical data for (ET)$_2$X salts (X=Cu(NCS)$_2$
[\onlinecite{a23}], Cu[N(CN)2]Br [\onlinecite{a24}] and Cu[N(CN)2]Cl
[\onlinecite{a25}]).  It does not seem particularly worthwhile to
determine the parameters of the Drude model in the case of
significantly different shape of optical conductivity as in ET salts
(see also [\onlinecite{a24,a25}]).

Eldridge et al. [\onlinecite{a24}] proposed to describe
$\sigma(\omega)$ for (ET)$_2$Cu[N(CN)$_2$]Br in terms of intra and
inter-band transitions. They used~\cite{a24} theoretical calculations
of the electronic structure, which had been performed at $T$=0 and not
taken electronic correlations into account. They could not explain
temperature evolution of spectra, which showed a decrease of the
intensity in low frequency range and increase of the intensity in the
charge transfer band. (ET)$_2$Cu[N(CN)$-2$]Br lies in the Kanoda phase
diagram (Fig.2) close to the Mott transition. It resembles the
situation studied by Pruschke et al. [\onlinecite{a26}] for the
Hubbard model of the infinite dimension. The electronic structure
shows two overlapping bands, if the system is very close to the Mott
transition from the metallic side. They found~\cite{a26}, in contrast
to the result obtained by Hubbard~\cite{a27}, an additional resonant
band exactly in the middle (the {\it Abrikosov-Suhl resonance}). The
amplitude of the resonant band is temperature dependent and increases
with $T\to 0$. The optical conductivity shows in this case the Drude
peak as well as the CT band, which temperature intensities are mutual
dependent. Since deuterated (ET)$_2$Cu[N(CN)$_2$]Br is exactly on the
border between the metal and the Mott insulator, one expects to see
the resonant band more pronounce (to our knowledge the experiment has
not been performed). The effective mass $m^*$ and other Fermi liquid
parameters should be extremely large in this case.

\section{Conclusions}

We have presented here a short review of physical properties of
molecular systems with correlated electrons. Main attention has been
focussed on a series of $\kappa$-(ET)$_2$X (X = Cu(NCS)$_2$,
Cu[N(CN)$_2$]Br and Cu[N(CN)$_2$]Cl). Their phase diagram is
qualitatively similar to that one for high temperature
superconductors. A comparison with the other systems of correlated
electrons showed that the effective mass $m^*$ in the ET salts is a
moderate value as well as the Sommerfeld coefficient $\gamma$. There
are no experimental data on an increase of $m^*$ when the system
approaches to the Mott transition from the metallic side (i.e. in the
salts with Cu(NCS)$_2$, Cu[N(CN)$_2$]Br and its deuterated form). The
Wilson ratio in these compounds is $R\approx$ 2 as in heavy fermions
(Fig.4). Precise measurements of resistivity and its anisotropy are
needed to determine the coefficient $A$. It would clarify the origin
of correlations; whether they are a coulombic or a phononic nature.
The photoemission data indicate on strong correlations~\cite{a14}, but
there is still lack of evidence for the Mott transition. The optical
conductivity measurements show~\cite{a23,a24} a flow of the spectral
weight from the Drude peak to the CT band with increasing a
temperature. One can assign it to changes of the intensity of the
Abrikosov-Suhl resonance for the system close to the Mott
transition.~\cite{a26}. We expect that an optical conductivity
experiment for deuterated (ET)$_2$Cu[N(CN)$_2$]Br should give more
evidence for the Mott transition.

In this lecture we did not presented properties of $d$=1 organic
compounds as (TMTTF)$_2$Y and (TMTSF)$_2$Y (with Y= PF$_6$, Br,
ClO$_4$), which have many interesting properties.~\cite{a28} For
example, the phase diagram~\cite{a28} has also the AF and
superconducting phases and is similar to that one in Fig.2. Correlated
electrons in the $d$=1 systems are analyzed within the Luttinger
liquid theory. They properties are different than those for the Fermi
liquid. The one-particle Green's function is given then by
\begin{equation}\label{e17}
G(x,t)\propto \exp(ik_Fx)\prod_{\alpha=\rho,\sigma}\frac{1}{[x-u_{\alpha}t+i/\tau]^{1/2+\eta_{\alpha}/2}[x+u_{\alpha}t-i/\tau]^{\eta_{\alpha}/2}}\;.
\end{equation}
There is a separation of charge $\rho$ and spin $\sigma$, which can
move with different velocities $u_{\rho}$ and $u_{\sigma}$.  The form
(\ref{e17}) is different than the Green's function (\ref{e5}) for the
Fermi liquid. The function (\ref{e17}) has no real poles, what means
that the picture of quasiparticles does not work in this case. Many
physical properties are therefore different.~\cite{a28}

\acknowledgments The Committee for Scientific Research (KBN) has
supported the work within the project No. 7T08A 003 12.

\newpage \epsfxsize=10cm \epsffile{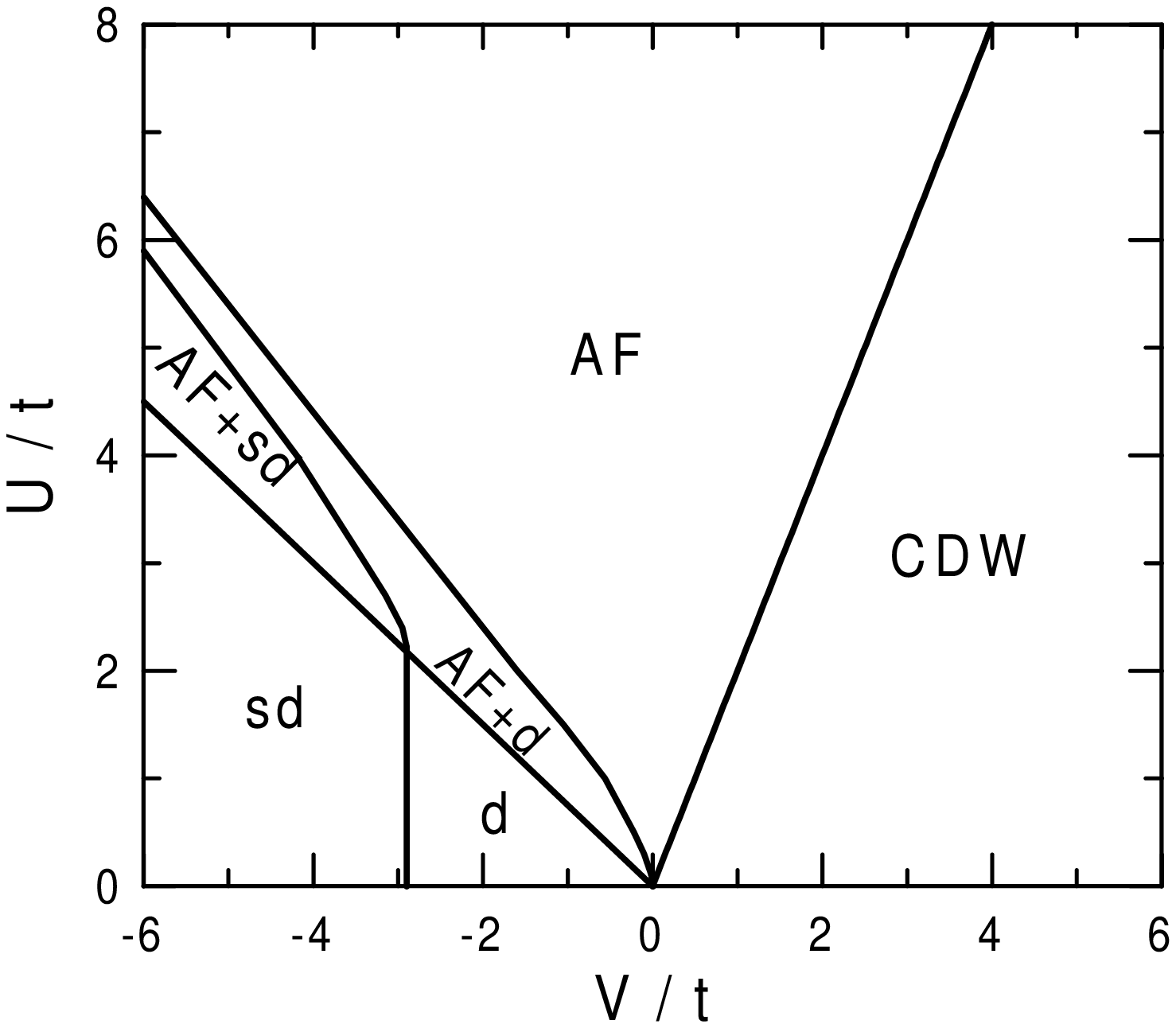} \hskip 1cm Fig.1 \vskip
0.1cm The phase diagram in the space of the parameters $U$ (onsite
interaction) and $V$ (intersite density-density interaction) of the
$d$=2 extended Hubbard model determined within the HF approximation
for the half-filled band (the electron concentration $n$=1) at
temperature $T$ = 0.  There are denoted the regions of the
antiferromagnetic (AF), the charge density wave (CDW) and the
superconducting state of the d-type (d) and the mixed s and d-type
pairing (sd).  Superconductivity and AF can coexist (in the region
denoted AF+sd and AF+d) (after [\onlinecite{a8}]).

\newpage
\epsfxsize=10cm
\epsffile{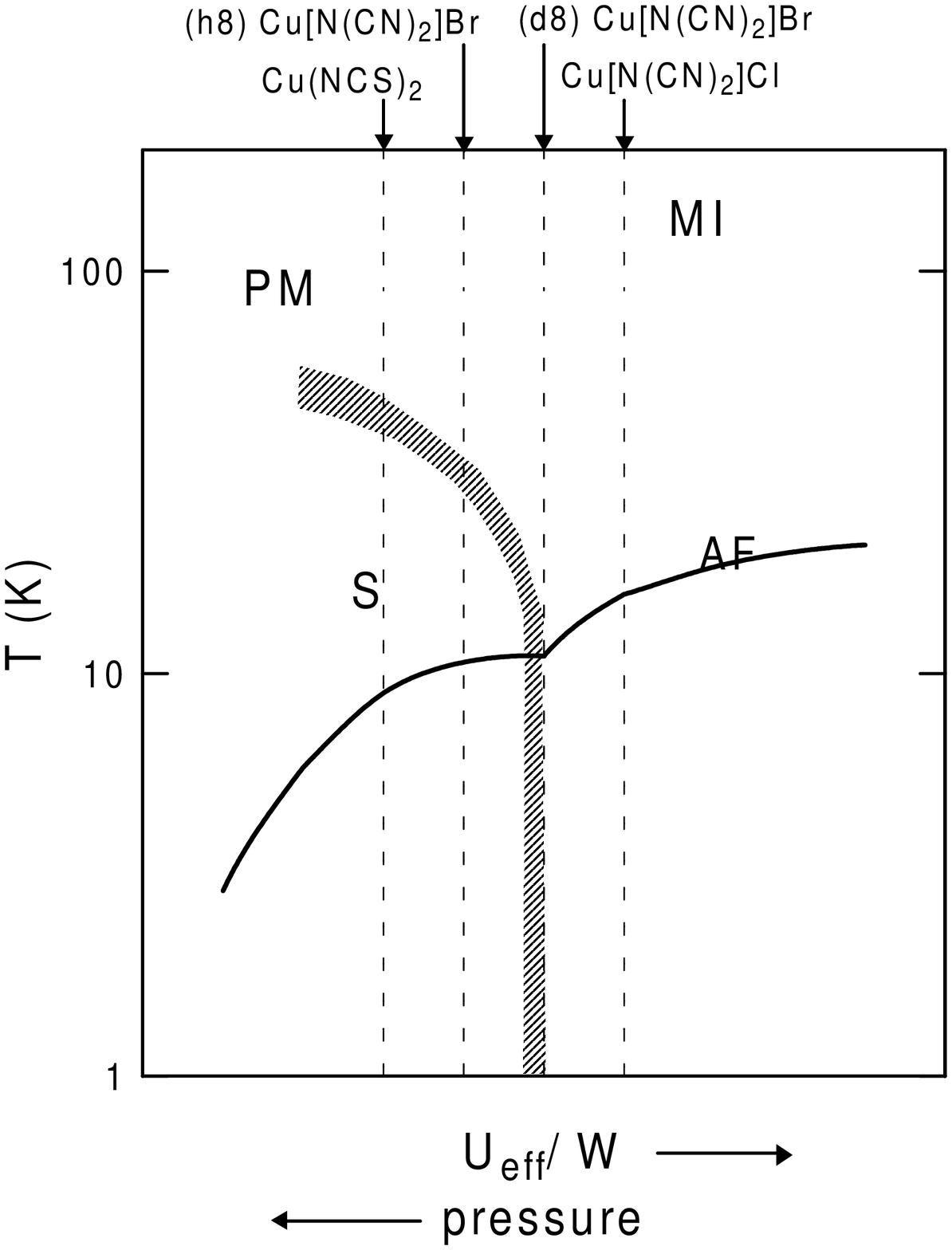 }
\hskip 1cm
Fig.2
\vskip 0.1cm
Schematic phase diagram for $\kappa$-(ET)$_2$X with X= Cu(NCS)$_2$,
Cu[N(CN)$_2$]Br, Cu[N(CN)$_2$]Cl and deuterated
(ET)$_2$Cu[N(CN)$_2$]Br denoted as (d8) Cu[N(CN)$_2$]Br. The regions
for the superconducting, the antiferromagnetic, the paramagnetic metal
as well as the Mott insulator phase are denoted as S, AF, PM and MI,
respectively.  (after [\onlinecite{a10}]).

\newpage
\epsfxsize=12cm
\epsffile{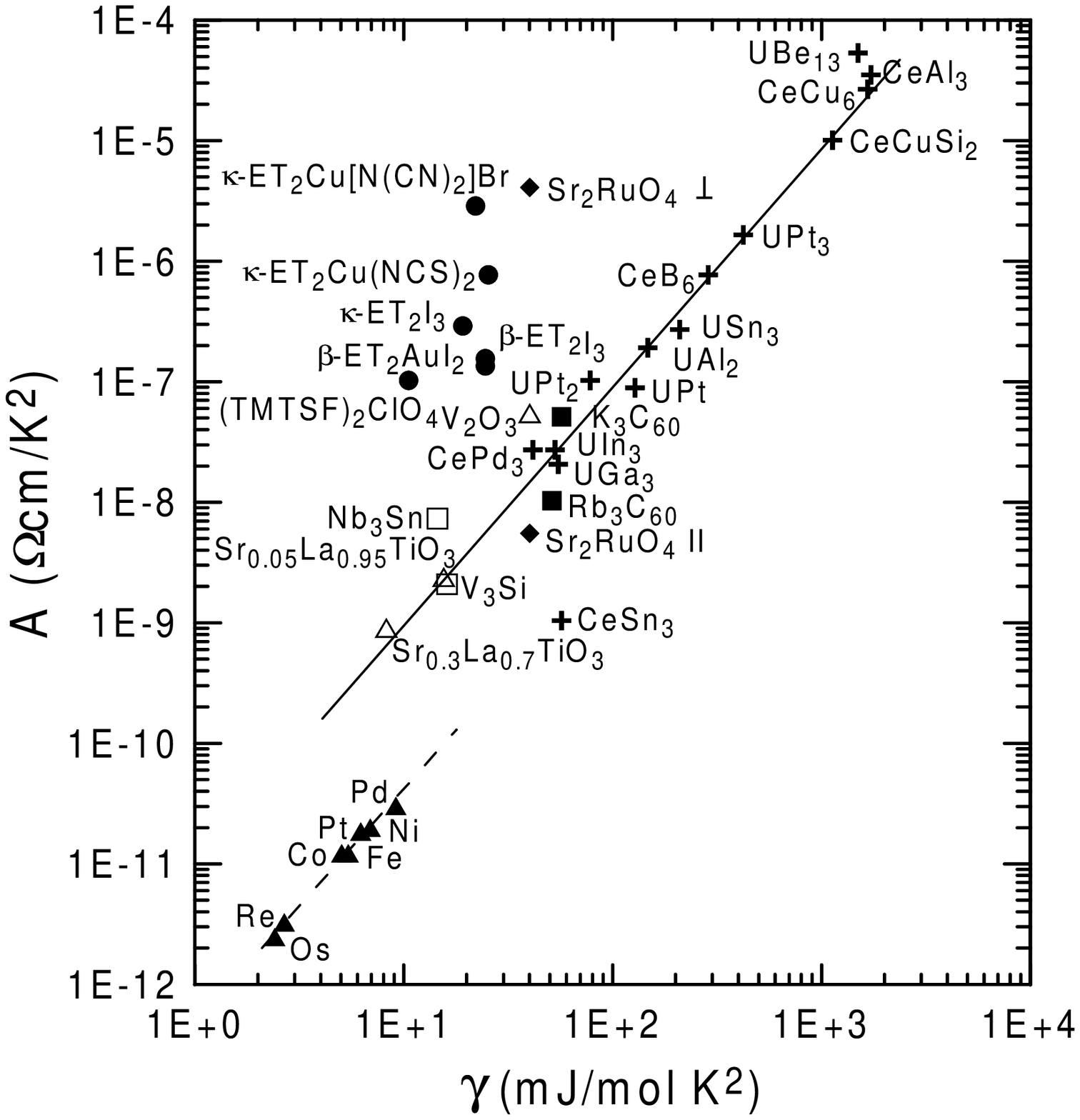 }
\hskip 1cm
Fig.3
\vskip 0.1cm
Resistivity coefficient $A$ plotted vs. the Sommerfeld coefficient
$\gamma$ for heavy fermion systems (crosses), fullerenes (full boxes),
A15 superconductors (open boxes), metal oxides (open triangles),
organic conductors (dots) and transition metals (full triangles)
(after [\onlinecite{a16,a17,a19}]). Diamonds denote the data for
Sr$_2$RuO$_4$ measured in a direction parallel ($\parallel$) and
perpendicular ($\perp$) to the conducting plane (after
[\onlinecite{a18}]).

\newpage
\vskip 12cm
\epsfxsize=12cm
\epsffile{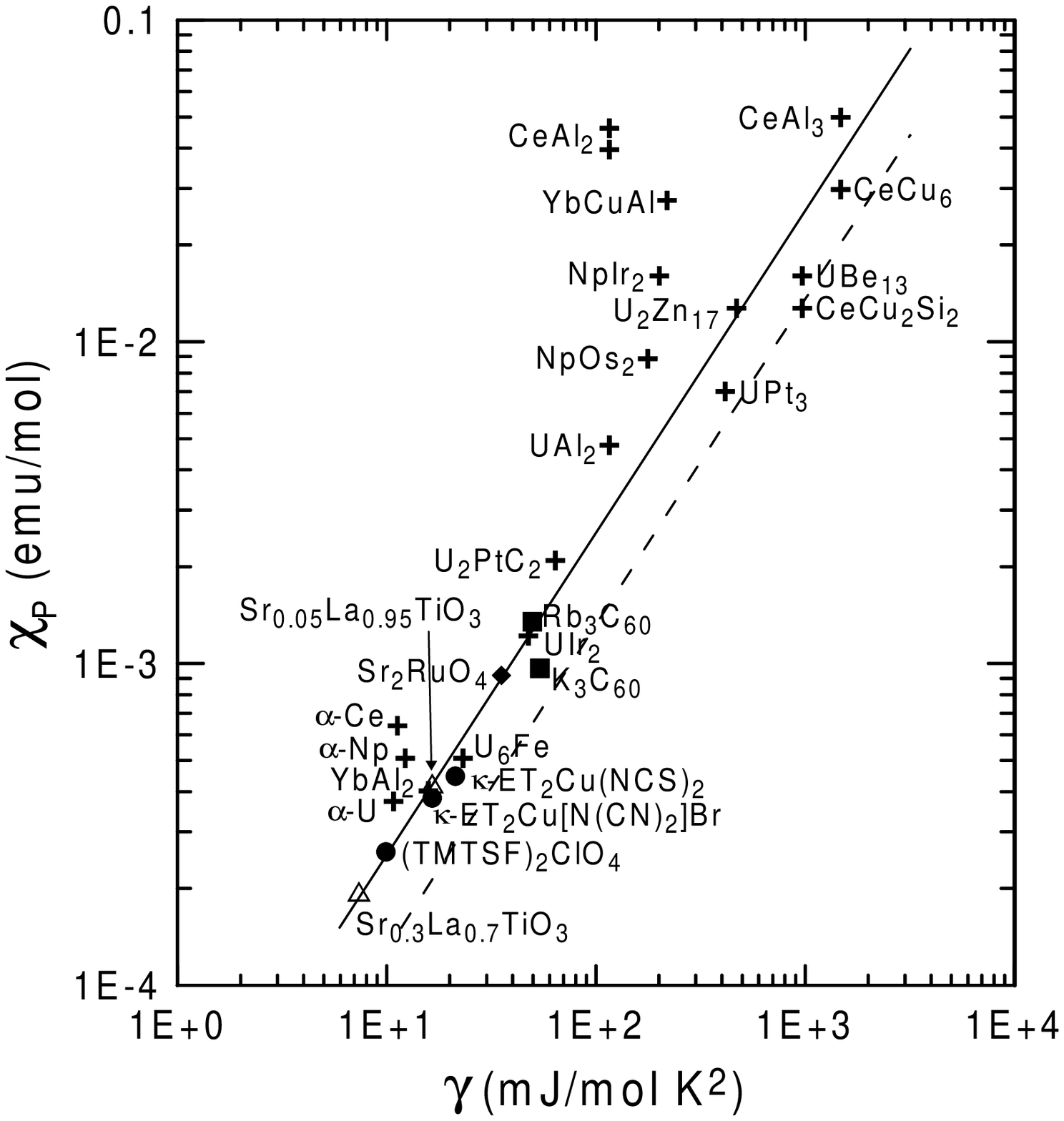 }
\hskip 1cm
Fig.4
\vskip 0.1cm
Pauli susceptibility plotted vs. the Sommerfeld coefficient $\gamma$
for heavy fermion systems (crosses) [\onlinecite{a18}], metal oxides
(open triangles) [\onlinecite{a19}] and organic conductors (dots).
Diamond denotes the point for Sr$_2$RuO$_4$ (after
[\onlinecite{a18}]). The solid and dashed lines correspond to the
Wilson ratio $R$ = 2 and 1, respectively.

\end{document}